\documentclass[conference, a4paper]{IEEEtran}

\usepackage{graphicx, setspace, acronym, ifthen, substr, color}
\usepackage[pdftitle={Community Cloud Computing}, pdfauthor={Gerard Briscoe, Alexandros Marinos}, pdfsubject={Community Cloud}, pdfkeywords={Cloud Computing, Community Cloud, Community Cloud Computing, Green Computing, Sustainability}, colorlinks=true, linkcolor=black, citecolor=black, filecolor=black, urlcolor=black, hyperfootnotes=false]{hyperref} 
\usepackage[UKenglish]{babel}
\usepackage[paper=a4paper,left=2cm,right=2cm,top=2cm,bottom=2cm]{geometry}
\usepackage[hang,splitrule]{footmisc}
\usepackage[normalem]{ulem}
\usepackage{breakurl}

\definecolor{tred}{RGB}{255,0,0}

\newboolean{final}\setboolean{final}{true}
\newboolean{arxiv}\setboolean{arxiv}{true}
\newboolean{mactex}\setboolean{mactex}{true}

\ifthenelse{\boolean{final}}{}{\usepackage{showlabels}\usepackage{citeext}}

\ifthenelse{\boolean{arxiv}}
	{\newcommand{\arxivfootnote}{\footnote}}
	{\def\arxivfootnote#1{}}

\ifthenelse{\boolean{arxiv}}
	{\def\arxivOnly#1{#1}}
	{\def\arxivOnly#1{}}

\ifthenelse{\boolean{arxiv}}
	{\def\arxivNot#1{}}
	{\def\arxivNot#1{#1}}

\ifthenelse{\boolean{arxiv}}
	{}
	{}

\newcommand{\tfigure}[9]
	{
	\IfSubStringInString{!}{#7}{\begin{figure}[#7]}{\begin{figure}[!t]}
	\IfSubStringInString{mm}{#8}{\vspace{#8}}{}
	\centering

	\IfSubStringInString{pdf}{#3}
		{
		\ifthenelse{\boolean{mactex}}{}{\execute{cd images; ln -s #2.pdf .#2.gdf}}
		\includegraphics[#1]{images/#2}
		}
		{
		\ifthenelse{\boolean{mactex}}{}{\execute{cd images; ./pdfcrop.sh #2}}
		\includegraphics[#1]{images/#2-crop.pdf}
		}
	\vspace{#6}
	\caption[#4]
		{
		\label{#2}
		#4: #5
		}
	\IfSubStringInString{mm}{#9}{\vspace{#9}}{}
	\end{figure}
	}

\newcommand{\Circlesub}[4]
	{
	\ifthenelse{\boolean{mactex}}{}{\immediate\write18{cd images; ./pdfcrop.sh circle#2}}
	\ifthenelse{\boolean{final}}
		{\hspace{#1}\raisebox{#4}{$\includegraphics[scale=0.5, clip=true, trim=0mm 0mm 0mm 0mm]{images/circle#2-crop.pdf}$}\hspace{#3}}
		{\href{file://localhost/Users/g/Desktop/PhDthesis/images/circle#2.graffle}{\hspace{#1}\raisebox{#4}{$\includegraphics[clip=true, trim=0mm 0mm 0mm 0mm]{images/circle#2-crop.pdf}$}\hspace{#3}}}
	}

\newcommand{\execute}[1]{\immediate\write18{#1}}

\newcommand{\white}[1]{\color{white}#1\normalcolor}

\newcommand{\setCap}[2]{#1\immediate\write18{./mkcaption.sh #2}}
\newcommand{\getCap}[1]{\acl*{#1}}

\acrodef{col3cap}{when consumers visit an application served by the central Cloud, which is housed in one or more data centres \cite{buyya2008market}.}
\acrodef{colCap}{{Green} symbolises resource {consumption}, and {yellow} resource {provision}}
\acrodef{col2Cap}{The role of {coordinator} for resource provision is designated by {red}, and is centrally controlled.}
\acrodef{absCap}{While there is a significant {buzz} around Cloud Computing, there is little clarity over which offerings qualify or their interrelation. The key to resolving this confusion is the realisation that the various offerings fall into different levels of abstraction,}
\acrodef{abs2Cap}{aimed at different market segments.}
\acrodef{c32cap}{shaping the underutilised resources of user machines}
\acrodef{c3cap}{with nodes potentially fulfilling all roles, {consumer}, {producer}, and most importantly {coordinator}}
\acrodef{c3CapArch}{most fundamental layer deals with distributing {coordination}}
\acrodef{c3Cap2Arch}{One layer above, {resource} provision and consumption are arranged on top of the coordination framework.}
\acrodef{c3Cap3Arch}{{service} layer is where resources are combined into end-user accessible services, to then themselves be composed into higher-level services.}

\acrodef{PCG}{Projected Conjugate Gradient} 
\acrodef{QP}{quadratic programming}
\acrodef{RBF}{Radial-Basis Function}
\acrodef{ABM}{Agent-Based Modelling}
\acrodef{AI}{Artificial Intelligence}
\acrodef{DAI}{Distributed Artificial Intelligence}
\acrodef{API}{Application Programming Interface}
\acrodef{ARF}{p14ARF human tumor-suppressor gene}
\acrodef{B2B}{business-to-business}
\acrodef{BDP}{Biological Design Pattern}
\acrodef{BGS}{Best Guess Solution}
\acrodef{BIC}{Biologically-Inspired Computing}
\acrodef{BML}{Business Modelling Language}
\acrodef{BPEL}{Business Process Execution Language}
\acrodef{BPMN}{Business Process Modelling Notation}
\acrodef{CAS}{Complex Adaptive Systems}
\acrodef{COBOL}{COmmon Business-Oriented Language}
\acrodef{DBE}{Digital Business Ecosystem}
\acrodef{DE}{Digital Ecosystem}
\acrodef{DEC}{distributed evolutionary computing}
\acrodef{DGA}{Distributed genetic algorithms}
\acrodef{DIS}{Distributed Intelligence System}
\acrodef{DNA}{Deoxyribose Nucleic Acid}
\acrodef{DOP}{DBE Open Protocol}
\acrodef{DSS}{Distributed Storage System}
\acrodef{EAP}{Evolving Agent Population}
\acrodef{ebXML}{e-business eXtensible Markup Language}
\acrodef{EC}{Evolutionary Computing}
\acrodef{ECJ}{Evolutionary Computing in Java}
\acrodef{EE}{Evolutionary Environment}
\acrodef{EFL}{Evolutionary Framework for Language}
\acrodef{FLE}{Framework for Language Ecosystems}
\acrodef{EOA}{Ecosystem-Oriented Architecture}
\acrodef{ESS}{evolutionary stable strategy}
\acrodef{EvE}{Evolutionary Environment}
\acrodef{ExE}{Execution Environment}
\acrodef{FCB}{Framework for Computational Biomimicry}
\acrodef{FFF}{Fitness Function Framework}
\acrodef{FL}{Fitness Landscape}
\acrodef{HWU}{Heriot-Watt University}
\acrodef{ICL}{Imperial College London}
\acrodef{ICT}{Information and Communications Technology}
\acrodef{INTEL}{Intel Ireland}
\acrodef{IPA}{International Phonetic Alphabet}
\acrodef{ISUFI}{Istituto Superiore Universitario di Formazione Interdisciplinare}
\acrodef{JDJ}{Java Developer's Journal}
\acrodef{KC}{Kolmogorov-Chaitin}
\acrodef{LAN}{local area network}
\acrodef{LSE}{London School of Economics and Political Science}
\acrodef{MAS}{Multi-Agent System}
\acrodef{MDL}{Minimum Description Length}
\acrodef{MDM2}{murine double minute 2}
\acrodef{MFT}{Mean Field Theory}
\acrodef{MoAS}{Mobile Agent System}
\acrodef{MOF}{Meta Object Facility}
\acrodef{MUH}{migration and usage history}
\acrodef{NIC}{Nature Inspired Computing}
\acrodef{NN}{Neural Network}
\acrodef{NoE}{Network of Excellence}
\acrodef{OMG}{Open Mac Grid}
\acrodef{OPAALS}{Open Philosophies for Associative Autopoietic Digital Ecosystems}
\acrodef{P2P}{peer-to-peer}
\acrodef{P53}{protein 53}
\acrodef{PDA}{Personal Digital Assistant}
\acrodef{QoS}{quality of service}
\acrodef{REST}{REpresentational State Transfer}
\acrodef{RNA}{Deoxyribose Nucleic Acid}
\acrodef{SAE}{Software Agent Ecosystem}
\acrodef{SBML}{Systems Biology Modelling Language}
\acrodef{SBVR}{Semantics of Business Vocabulary and Business Rules}
\acrodef{SDL}{Service Description Language}
\acrodef{SF}{Service Factory}
\acrodef{SIM}{Social Interaction Mechanism}
\acrodef{SM}{Service Manifest}
\acrodef{SME}{Small and Medium sized Enterprise}
\acrodef{SML}{Service Modelling Language}
\acrodef{SMO}{Sequential Minimal Optimisation}
\acrodef{SOA}{Service-Oriented Architecture}
\acrodef{SOAP}{Simple Object Access Protocol}
\acrodef{SOC}{Self-Organised Criticality}
\acrodef{SOLUTA}{SOLUTA.NET}
\acrodef{SOM}{Self-Organising Map}
\acrodef{SSL}{Semantic Service Language}
\acrodef{STU}{Salzburg Technical University}
\acrodef{SUN}{Sun Microsystems}
\acrodef{SVM}{Support Vector Machine}
\acrodef{TM}{Turing Machine}
\acrodef{UBHAM}{University of Birmingham}
\acrodef{UDDI}{Universal Description Discovery and Integration}
\acrodef{UML}{Unified Modelling Language}
\acrodef{URI}{Uniform Resource Identifier}
\acrodef{UTM}{Universal Turing Machine}
\acrodef{VLP}{variable length population}
\acrodef{VLS}{variable length sequences}
\acrodef{vls}{variable length sequence}
\acrodef{WP}{Work-Package}
\acrodef{WSDL}{Web Services Definition Language}
\acrodef{XMI}{XML Metadata Interchange}
\acrodef{XML}{eXtensible Markup Language}
\acrodef{MD5}{Message-Digest algorithm 5}
\acrodef{GA}{genetic algorithm}
\acrodef{GP}{genetic programming}
\acrodef{MASON}{Multi-Agent Simulator Of Neighbourhoods}
\acrodef{Repast}{Recursive Porous Agent Simulation Toolkit}
\acrodef{JCLEC}{Java Computing Library for Evolutionary Computing}
\acrodef{OWL-S}{Web Ontology Language - Service}
\acrodef{EGT}{Evolutionary Game Theory}
\acrodef{RBF}{Radial Basis Functions}
\acrodef{SWS}{Semantic Web Services}
\acrodef{HDD}{Hard Disk Drive}
\acrodef{SSD}{Solid-State Drive}
\acrodef{OKS}{Open Knowledge Space}
\acrodef{CAES}{Complex Adaptive EcoSystem}
\acrodef{SaaS}{Software-as-a-Service}
\acrodef{PaaS}{Platform-as-a-Service}
\acrodef{IaaS}{Infrastructure-as-a-Service}
\acrodef{C3}{Community Cloud Computing}

\begin{document}

\title{Community Cloud Computing}
\author{
 \IEEEauthorblockN{Alexandros Marinos}
 \IEEEauthorblockA{Department of Computing\\
University of Surrey\\
United Kingdom\\
e-mail: a.marinos@surrey.ac.uk}
\and 
 \IEEEauthorblockN{Gerard Briscoe}
 \IEEEauthorblockA{Department of Media and Communications\\
London School of Economics and Political Science\\
United Kingdom\\
e-mail: g.briscoe@lse.ac.uk}
}
\maketitle

\begin{abstract}
\white{.} Cloud Computing is rising fast, with its data centres growing at an unprecedented rate. However, this has come with concerns over privacy, efficiency at the expense of resilience, and environmental sustainability, because of the dependence on Cloud vendors such as Google, Amazon and Microsoft. Our response is an alternative model for the Cloud conceptualisation, providing a paradigm for Clouds in the community, utilising networked personal computers for liberation from the centralised vendor model. \ac{C3} offers an alternative architecture, created by combing the Cloud with paradigms from Grid Computing, principles from Digital Ecosystems, and sustainability from Green Computing, while remaining true to the original vision of the Internet. It is more technically challenging than Cloud Computing, having to deal with distributed computing issues, including heterogeneous nodes, varying quality of service, and additional security constraints. However, these are not insurmountable challenges, and with the need to retain control over our digital lives and the potential environmental consequences, it is a challenge we must pursue.
\end{abstract}

\section{Introduction}

The recent development of Cloud Computing provides a compelling value proposition for organisations to outsource their \ac{ICT} infrastructure \cite{wiki1}. However, there are growing concerns over the control ceded to large Cloud vendors \cite{wrongCloud}, especially the lack of information privacy \cite{berkely}. Also, the data centres required for Cloud Computing are growing exponentially \cite{hayes}, creating an ever-increasing \emph{carbon footprint} and therefore raising environmental concerns \cite{mckenna2008cws, mckinsey}.

The distributed resource provision from Grid Computing, distributed control from Digital Ecosystems, and sustainability from Green Computing, can remedy these concerns. So, Cloud Computing combined with these approaches would provide a compelling socio-technical conceptualisation for sustainable distributed computing, utilising the spare resources of networked personal computers collectively to provide the facilities of a virtual \emph{data centre} and form a Community Cloud. Therefore, essentially reformulating the Internet to reflect its current uses and scale, while maintaining the original intentions \cite{leiner2001bhi} for sustainability in the face of adversity. Including extra capabilities embedded into the infrastructure which would become as fundamental and invisible as moving packets is today.

\section{Cloud Computing}

Cloud Computing is the use of Internet-based technologies for the provision of services \cite{wiki1}, originating from the \emph{cloud} as a metaphor for the Internet, based on depictions in computer network diagrams to abstract the complex infrastructure it conceals \cite{wiki7}. It can also be seen as a commercial evolution of the academic-oriented Grid Computing \cite{foster2008cca}, succeeding where Utility Computing struggled \cite{utilityComp, cellchip}, while making greater use of the self-management advances of Autonomic Computing \cite{kephart2003vac}. It offers the illusion of infinite computing resources available on demand, with the elimination of upfront commitment from users, and payment for the use of computing resources on a short-term basis as needed \cite{berkely}. Furthermore, it does not require the node providing a service to be present once its service is deployed \cite{berkely}. It is being promoted as the cutting-edge of scalable web application development \cite{berkely}, in which dynamically scalable and often virtualised resources are provided as a service over the Internet \cite{wiki2, wiki1, wiki4, wiki5}, with users having no knowledge of, expertise in, or control over the technology infrastructure of the Cloud supporting them \cite{wiki6}. It currently has significant momentum in two extremes of the web development industry \cite{berkely, wiki1}: the consumer web technology incumbents who have resource surpluses in their vast \emph{data centres}\arxivfootnote{A \emph{data centre} is a facility, with the necessary security devices and environmental systems (e.g. air conditioning and fire suppression), for housing a \emph{server farm}, a collection of computer servers that can accomplish server needs far beyond the capability of one machine \cite{arregoces2003dcf}.}, and various consumers and start-ups that do not have access to such computational resources. Cloud Computing conceptually incorporates \ac{SaaS} \cite{saas}, Web 2.0 \cite{web2.0} and other technologies with reliance on the Internet, providing common business applications online through web browsers to satisfy the computing needs of users, while the software and data are stored on the servers.

\tfigure{scale=2}{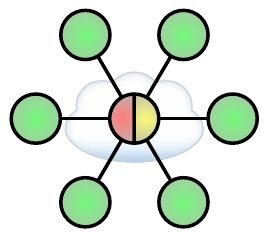}{graffle}{Cloud Computing}{Typical configuration \getCap{col3cap} \getCap{colCap}. \getCap{col2Cap}}{-2mm}{!h}{}{}

Figure \ref{cloudComputing} shows the typical configuration of Cloud Computing at run-time \setCap{when consumers visit an application served by the central Cloud, which is housed in one or more data centres \cite{buyya2008market}.}{col3cap} \setCap{\emph{Green} symbolises resource \emph{consumption}, and \emph{yellow} resource \emph{provision}}{colCap}. \setCap{The role of \emph{coordinator} for resource provision is designated by \emph{red}, and is centrally controlled.}{col2Cap} Even if the central node is implemented as a distributed grid, which is the usual incarnation of a data centre, control is still centralised. Providers, who are the controllers, are usually companies with other web activities that require large computing resources, and in their efforts to scale their primary businesses have gained considerable expertise and hardware. For them, Cloud Computing is a way to resell these as a new product while expanding into a new market. Consumers include everyday users, \acp{SME}, and ambitious start-ups whose innovation potentially threatens the incumbent providers.

\subsection{Layers of Abstraction}

\setCap{While there is a significant \emph{buzz} around Cloud Computing, there is little clarity over which offerings qualify or their interrelation. The key to resolving this confusion is the realisation that the various offerings fall into different levels of abstraction,}{absCap} as shown in Figure \ref{Drawing1}, \setCap{aimed at different market segments.}{abs2Cap}

\subsubsection{\ac{IaaS} \cite{iaas}} At the most basic level of Cloud Computing offerings, there are providers such as Amazon \cite{amazon} and Mosso \cite{mosso}, who provide \emph{machine instances} to developers. These instances essentially behave like dedicated servers that are controlled by the developers, who therefore have full responsibility for their operation. So, once a machine reaches its performance limits, the developers have to manually instantiate another machine and scale their application out to it. This service is intended for developers who can write arbitrary software on top of the infrastructure with only small compromises in their development methodology. 

\subsubsection{\ac{PaaS} \cite{paas}} One level of abstraction above, services like Google App Engine \cite{appEngine} provide a programming environment that abstracts machine instances and other technical details from developers. The programs are executed over data centres, not concerning the developers with matters of allocation. In exchange for this, the developers have to handle some constraints that the environment imposes on their application design, for example the use of \emph{key-value stores}\arxivfootnote{A distributed storage system for structured data that focuses on scalability, at the expense of the other benefits of relational databases \cite{keyvaluestore}, e.g. Google's BigTable \cite{bigtable} and Amazon's SimpleDB \cite{dynamo}.} instead of \emph{relational databases}.

\subsubsection{\acf{SaaS} \cite{saas}} At the consumer-facing level are the most popular examples of Cloud Computing, with well-defined applications offering users online resources and storage. This differentiates \ac{SaaS} from traditional websites or web applications which do not interface with user information (e.g. documents) or do so in a limited manner. Popular examples include Microsoft's (Windows Live) Hotmail, office suites such as Google Docs\arxivOnly{ and Zoho}, and online business software such as Salesforce.com.

To better understand Cloud Computing we can categorise the roles of the various actors. The \emph{vendor} as resource provider has already been discussed. The application \emph{developers} utilise the resources provided, building services for the \emph{end users}. This separation of roles helps define the stakeholders and their differing interests. However, actors can take on multiple roles, with \emph{vendors} also developing services for the \emph{end users}, or \emph{developers} utilising the services of others to build their own services. Yet, within each Cloud the role of provider, and therefore controller, can only be occupied by the \emph{vendor} providing the Cloud.

\tfigure{width=3.25in}{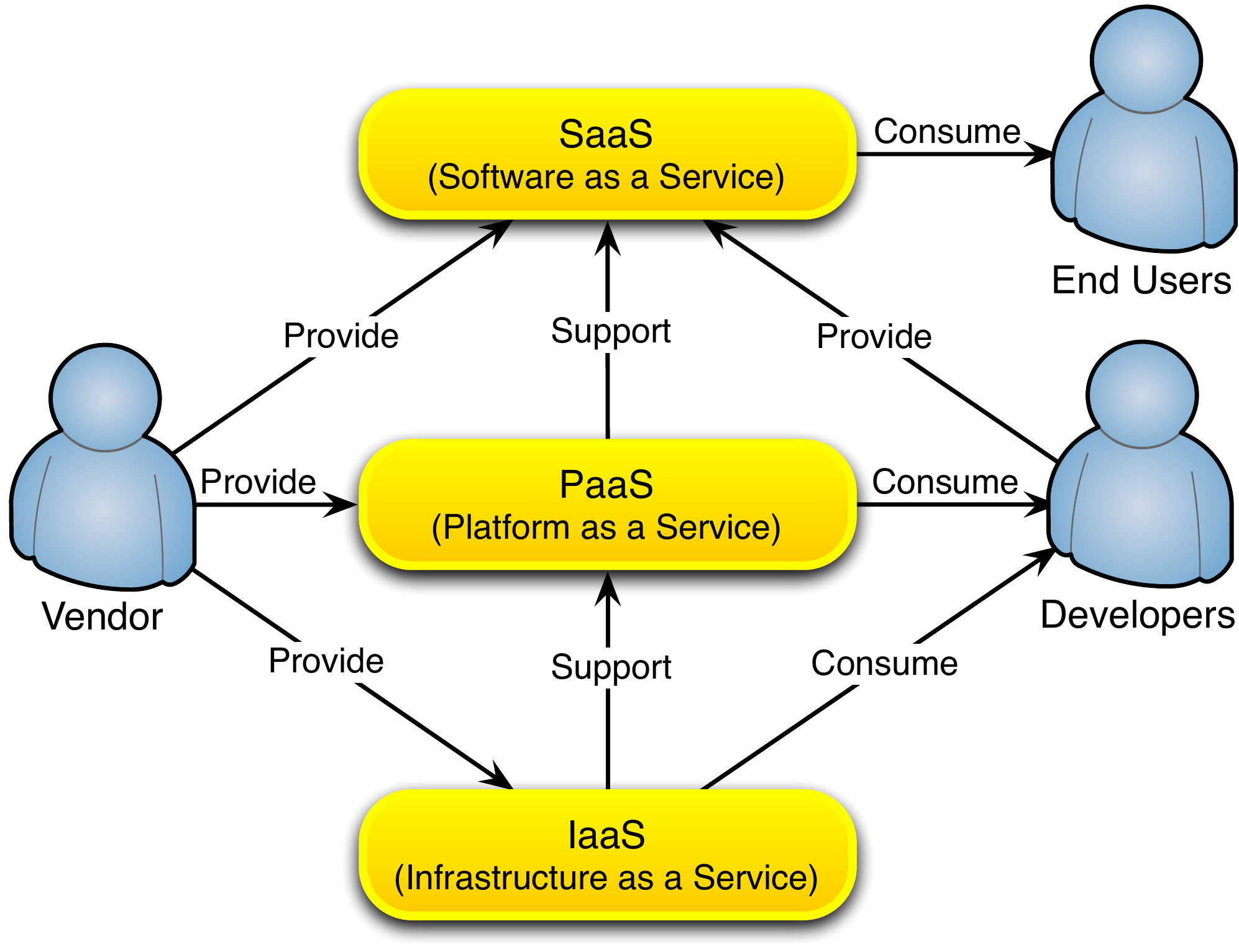}{graffle}{Abstractions of Cloud Computing}{\getCap{absCap} \getCap{abs2Cap}}{-7mm}{}{}{}

\subsection{Concerns}
The Cloud Computing model is not without concerns, as others have noted \cite{gnuMan, berkely}, and we consider the following as primary:

\subsubsection{Failure of Monocultures} The uptime\arxivfootnote{Uptime is a measure of the time a computer system has been running, i.e. up. It came into use to describe the opposite of downtime, times when a system was not operational \cite{mccabe2007naa}.} of Cloud Computing based solutions is an advantage, when compared to businesses running their own infrastructure, but often overlooked is the co-occurrence of downtime in vendor-driven \emph{monocultures}. The use of globally decentralised \emph{data centres} for vendor Clouds minimises failure, aiding its adoption. However, when a cloud fails, there is a cascade effect crippling all organisations dependent on that Cloud, and all those dependent upon them. This was illustrated by the Amazon (S3) Cloud outage \cite{amazonOutage}, which disabled several other dependent businesses. So, failures are now system-wide, instead of being partial or localised. Therefore, the efficiencies gained from centralising infrastructure for Cloud Computing are increasingly at the expense of the Internet's resilience.

\subsubsection{Convenience vs Control} The growing popularity of Cloud Computing comes from its convenience, but also brings vendor control, an issue of ever-increasing concern. For example, Google Apps for in-house e-mail typically provides higher uptime \cite{montgomery2008}, but its failure \cite{perez2007} highlights the issue of lock-in that comes from depending on vendor Clouds. The even greater concern is the loss of information privacy, with vendors having full access to the resources stored on their Clouds. So much so the British government is considering a `G Cloud' for government business applications \cite{gcloud}. In particularly sensitive cases of \acp{SME} and start-ups, the provider-consumer relationship that Cloud Computing fosters between the owners of resources and their users could potentially be detrimental, as there is a potential conflict of interest for the providers. They profit by providing resources to up-and-coming players, but also wish to maintain dominant positions in their consumer-facing industries.

\subsubsection{Environmental Impact} The other major concern is the ever-increasing \emph{carbon footprint} from the \emph{exponential growth} \cite{hayes} of the \emph{data centres} required for Cloud Computing. With the industry expected to exceed the airline industry by 2020 \cite{mckinsey}, raising sustainability concerns \cite{mckenna2008cws}. The industry is being motivated to address the problem by legislation \cite{mckinsey, epaReport}, the operational limit of power grids (being unable to power anymore servers in their data centres) \cite{miller2006}, and the potential financial benefits of increased efficiency \cite{mcIsaac2007, mckinsey}. Their primary solution is the use of \emph{virtualisation}\arxivfootnote{Virtualisation is the creation of a virtual version of a resource, such as a server, which can then be stored, migrated, duplicated, and instantiated as needed, improving scalability and work load management \cite{wolf2005vde}.} to maximise resource utilisation \cite{virtualisation}, but the problem remains \cite{brill2007icd, brodkin2008}.

While these issues are endemic to Cloud Computing, they are not flaws in the Cloud conceptualisation, but the vendor provision and implementation of Clouds \cite{appEngine, amazon, azure}. There are attempts to address some of these concerns, such as a portability layer between vendor Clouds to avoid lock-in \cite{metaCloud}. However, this will not alleviate issues such as inter-Cloud latency \cite{latency}. An open source implementation of the Amazon (EC2) Cloud \cite{amazon}, called Eucalyptus \cite{nurmi2008eos}, allows a data centre to execute code compatible with Amazon's Cloud. Allowing for the creation of \emph{private internal} Clouds, avoiding vendor lock-in and providing information privacy, but only for those with their own data centre and so is not really Cloud Computing (which by definition is to avoid owning data centres \cite{wiki1}). Therefore, vendor Clouds remain synonymous with Cloud Computing \cite{wiki2, wiki1, wiki4, wiki5}. Our response is an alternative model for the Cloud conceptualisation, created by combining the Cloud with paradigms from Grid Computing, principles from Digital Ecosystems, and sustainability from Green Computing, while remaining true to the original vision of the Internet \cite{abbate1999ii}.

\section{Grid Computing: Distributing Provision}
Grid Computing is a form of distributed computing in which a \emph{virtual super computer} is composed from a cluster of networked, loosely coupled computers, acting in concert to perform very large tasks \cite{foster2004gbn}. It has been applied to computationally intensive scientific, mathematical, and academic problems through volunteer computing, and used in commercial enterprise for such diverse applications as drug discovery, economic forecasting, seismic analysis, and back-office processing to support e-commerce and web services \cite{foster2004gbn}.

\tfigure{scale=2}{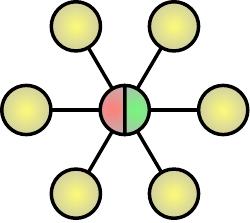}{graffle}{Grid Computing}{Typical configuration in which resource provision is managed by a group of distributed nodes \cite{foster2004gbn}. \getCap{colCap}. \getCap{col2Cap}}{-2mm}{}{}{}

What distinguishes Grid Computing from \emph{cluster computing} is being more loosely coupled, heterogeneous, and geographically dispersed \cite{foster2004gbn}. Also, grids are often constructed with general-purpose grid software libraries and middleware, dividing and apportioning pieces of a program to potentially thousands of computers \cite{foster2004gbn}. However, what distinguishes Cloud Computing from Grid Computing is being web-centric, despite some of its definitions being conceptually similar (such as computing resources being consumed as electricity is from power grids) \cite{foster2008cca}.

\section{Digital Ecosystems: Distributing Control}

Digital Ecosystems are distributed adaptive open socio-technical systems, with properties of self-organisation, scalability and sustainability, inspired by natural ecosystems \cite{bionetics, thesis}. Emerging as a novel approach to the catalysis of sustainable regional development driven by \acp{SME} \cite{reden}. Aiming to help local economic actors become active players in globalisation \cite{dini2008bid}, valorising their local culture and vocations, and enabling them to interact and create value networks at the global level \cite{dbebook}. Increasingly this approach, dubbed \emph{glocalisation}, is being considered a successful strategy of globalisation that preserves regional growth and identity \cite{robertson1994gog, \arxivOnly{swyngedouw1992mqg,} khondker2004}, and has been embraced by the mayors and decision-makers of thousands of municipalities \cite{glocalforum2004}. The community focused on the deployment of Digital Ecosystems, REgions for Digital Ecosystems Network (REDEN) \cite{reden}, is supported by projects such as the Digital Ecosystems Network of regions for (4) DissEmination and Knowledge Deployment (DEN4DEK) \cite{den4dek}. This thematic network that aims to share experiences and disseminate knowledge to let regions effectively deploy of Digital Ecosystems at all levels (economic, social, technical and political) to produce real impacts in the economic activities of European regions through the improvement of \ac{SME} business environments.

In a traditional market-based economy, made up of sellers and buyers, the parties exchange property, while in a new network-based economy, made up of servers and clients, the parties share access to services and experiences \cite{delcloque2001dii}. Digital Ecosystems aim to support network-based economies reliant on next-generation \ac{ICT} that will extend the \ac{SOA} concept \cite{soa1w} with the automatic combining of available and applicable services in a scalable architecture, to meet business user requests for applications that facilitate business processes. Digital Ecosystems research is yet to consider scalable resource provision, and therefore risks being subsumed into vendor Clouds at the infrastructure level, while striving for decentralisation at the service level. So, the realisation of their vision requires a form of Cloud Computing, but with their principle of community-based infrastructure where individual users share ownership \cite{bionetics}.

\section{Green Computing: Growing Sustainably}
Green Computing is the efficient use of computing resources, with the primary objective being to account for the \emph{triple bottom line}\footnote{The triple bottom line (\emph{people, planet, profit}) \cite{elkington1998cft}.}, an expanded spectrum of values and criteria for measuring organisational (and societal) success \cite{williams2008guest}. Given computing systems existed before concern over their environmental impact, it has generally been implemented retroactively, but is now being considered at the development phase \cite{williams2008guest}. It is systemic in nature, because ever-increasingly sophisticated modern computer systems rely upon people, networks and hardware. So, the elements of a \emph{green} solution may comprise items such as end user satisfaction, management restructuring, regulatory compliance, disposal of electronic waste, telecommuting, virtualisation of server resources, energy use, thin client solutions and return on investment \cite{williams2008guest}.

One of the greatest environmental concerns of the industry is their data centres \cite{brodkin2008}, which have increased in number over time as business demands have increased, with facilities housing a rising amount of evermore powerful equipment \cite{arregoces2003dcf}. As data centres run into limits related to power, cooling and space, their ever-increasing operation has created a noticeable impact on power grids \cite{miller2006}. To the extent that data centre efficiency has become an important global issue, leading to the creation of the Green Grid \cite{greenGrid}, an international non-profit organisation mandating an increase in the energy efficiency of data centres. Their approach, virtualisation, has improved efficiency \cite{brill2007icd, brodkin2008}, but is optimising a flawed model that does not consider the whole system, where resource provision is disconnected from resource consumption. For example, competing vendors must host significant redundancy in their data centres to manage usage spikes and maintain the illusion of infinite resources. So, we would argue that an alternative more systemic approach is required, where resource consumption and provision are connected, to minimise the environmental impact and allow sustainable growth.

\section{Community Cloud}

\ac{C3} arises from concerns over Cloud Computing, specifically control by vendors and lack of environmental sustainability. The Community Cloud aspires to combine distributed resource provision from Grid Computing, distributed control from Digital Ecosystems and sustainability from Green Computing, with the use cases of Cloud Computing, while making greater use of self-management advances from Autonomic Computing. Replacing vendor Clouds by \setCap{shaping the underutilised resources of user machines}{c32cap} to form a Community Cloud, \setCap{with nodes potentially fulfilling all roles, \emph{consumer}, \emph{producer}, and most importantly \emph{coordinator}}{c3cap}, as shown in Figure \ref{c3}.

\tfigure{width=3.3in}{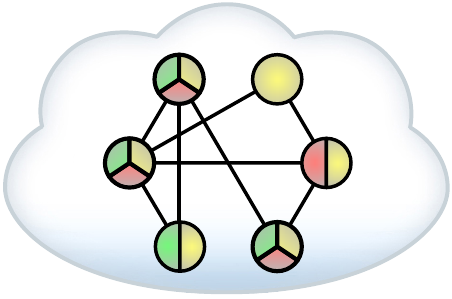}{graffle}{Community Cloud}{Created from \getCap{c32cap}, \getCap{c3cap}. Green symbolises resource consumption, yellow resource provision, and red resource coordination.}{-6mm}{!h}{}{}

\subsection{Conceptualisation}
The conceptualisation of the Community Cloud draws upon Cloud Computing \cite{buyya2008market}, Grid Computing \cite{foster2008cca}, Digital Ecosystems \cite{bionetics}, Green Computing \cite{harris2008} and Autonomic Computing \cite{kephart2003vac}. A paradigm for Cloud Computing in the \emph{community}, without dependence on Cloud vendors, such as Google, Amazon, or Microsoft.

\subsubsection{Openness}
Removing dependence on vendors makes the Community Cloud the open equivalent to vendor Clouds, and therefore identifies a new dimension in the open versus proprietary struggle \cite{west2001pvo} that has emerged in code, standards and data, but has yet to be expressed in the realm of hosted services. 
 
\subsubsection{Community}
The Community Cloud is as much a social structure as a technology paradigm \cite{benkler2004sns}, because of the community ownership of the infrastructure. Carrying with it a degree of economic scalability, without which there would be diminished competition and potential stifling of innovation as risked in vendor Clouds.

\subsubsection{Individual Autonomy}
In the Community Cloud, nodes have their own utility functions in contrast with data centres, in which dedicated machines execute software as instructed. So, with nodes expected to act in their own self-interest, centralised control would be impractical, as with consumer electronics like game consoles \cite{grand2004game}. Attempts to control user machines counter to their self-interest results in cracked systems, from black market hardware modifications and \emph{arms races} over hacking and securing the software (routinely lost by the vendors) \cite{grand2004game}. In the Community Cloud, where no concrete vendors exist, it is even more important to avoid antagonising the users, instead embracing their self interest and harnessing it for the benefit of the community with measures such as a \emph{community currency}.

\subsubsection{Identity}
In the Community Cloud each user would inherently possess a unique identity, which combined with the structure of the Community Cloud should lead to an inversion of the currently predominant membership model. So, instead of users registering for each website (or service) anew, they could simply add the website to their identity and grant access. Allowing users to have multiple services connected to their identity, instead of creating new identities for each service. This relationship is reminiscent of recent application platforms, such as Facebook's f8 and Apple's App Store, but decentralised in nature and so free from vendor control. Also, allowing for the reuse of the connections between users, akin to Google's Friend Connect, instead of reestablishing them for each new application.

\subsubsection{Graceful Failures} The Community Cloud is not owned or controlled by any one organisation, and therefore not dependent on the lifespan or failure of any one organisation. It therefore ought be robust and resilient to failure, and immune to the system-wide cascade failures of vendor Clouds, because of the diversity of its supporting nodes. When occasionally failing doing so gracefully, non-destructively, and with minimal downtime, as the unaffected nodes mobilise to compensate for the failure. 

\subsubsection{Convenience and Control} 
The Community Cloud, unlike vendor Clouds, has no inherent conflict between convenience and control, resulting from its community ownership providing distributed control, which would be more democratic. However, whether the Community Cloud can provide technically quality equivalent or superior to its centralised counterparts is an issue that will require further research.

\subsubsection{Community Currency}
The Community Cloud would require its own currency to support the sharing of resources, a \emph{community currency}, which in economics is a medium (currency), not backed by a central authority (e.g. national government), for exchanging goods and services within a community \cite{greco2001mua}. It does not need to be restricted geographically, despite sometimes being called a local currency \cite{doteuchi2002cca}. An example is the Fureai kippu system in Japan, which issues credits in exchange for assistance to senior citizens \cite{lietaer2004ccj}. Family members living far from their parents can earn credits by offering assistance to the elderly in their local community, which can then be transferred to their parents and redeemed by them for local assistance \cite{lietaer2004ccj}.

\subsubsection{Quality of Service} 
Ensuring acceptable \ac{QoS} in a heterogeneous system will be a challenge. Not least because achieving and maintaining the different aspects of \ac{QoS} will require reaching \emph{critical mass} in the participating nodes and available services. Thankfully, the \emph{community currency} could support long-term promises by resource providers and allow the higher quality providers, through market forces, to command a higher price for their service provision. Interestingly, the Community Cloud could provide a better \ac{QoS} than vendor Clouds, utilising time-based and geographical variations advantageously in the dynamic scaling of resource provision.

\subsubsection{Environmental Sustainability}
We expect the Community Cloud to have a smaller \emph{carbon footprint} than vendor Clouds, on the assumption that making use of underutilised user machines requires less energy than the dedicated data centres required for vendor Clouds. The server farms within data centres are an intensive form of computing resource provision, while the Community Cloud is more organic, growing and shrinking in a symbiotic relationship to support the demands of the community, which in turn supports it.

\subsubsection{Service Composition}

The great promise of service-oriented computing is that the \emph{marginal cost} of creating the n-th application will be virtually zero, as all the software required already exists to satisfy the requirements of other applications. Only their composition and orchestration are required to produce a new application \cite{tang2004ews, modi2008}. Within vendor Clouds it is possible to make services that expose themselves for composition and compose these services, allowing the hosting of a complete service-oriented architecture \cite{buyya2008market}. However, current service composition technologies have not gained widespread adoption \cite{SOAstandards}. Digital Ecosystems advocate service composability to avoid centralised control by large service providers, because easy service composition allows coalitions of \acp{SME} to compete simply by composing simpler services into more complex services that only large enterprises would otherwise be able to deliver \cite{dbebook}. So, we should extend decentralisation beyond resource provision and up to the service layer, to enable service composition within the Community Cloud.

\subsection{Architecture}

\tfigure{width=3.25in}{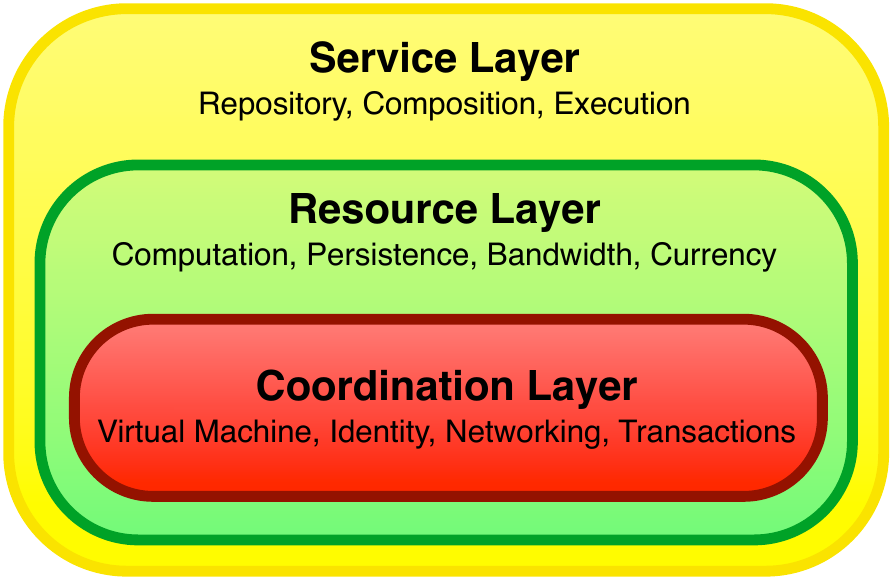}{graffle}{Community Cloud Computing}{An architecture in which the \getCap{c3CapArch}. \getCap{c3Cap2Arch} Finally, the \getCap{c3Cap3Arch}}{-6mm}{!h}{}{}

The method of materialising the Community Cloud is the distribution of its server functionality amongst a population of nodes provided by user machines, shaping their underutilised resources into a \emph{virtual data centre}. While straightforward in principle, it poses challenges on many different levels. So, an architecture for \ac{C3} can be divided into three layers, dealing with these challenges iteratively. The \setCap{most fundamental layer deals with distributing \emph{coordination}}{c3CapArch}, which is taken for granted in homogeneous data centres where good connectivity, constant presence and centralised infrastructure can be assumed. \setCap{One layer above, \emph{resource} provision and consumption are arranged on top of the coordination framework.}{c3Cap2Arch} Easy in the homogeneous grid of a data centre where all nodes have the same interests, but more challenging in a distributed heterogeneous environment. Finally, the \setCap{\emph{service} layer is where resources are combined into end-user accessible services, to then themselves be composed into higher-level services.}{c3Cap3Arch}

\subsubsection{Coordination Layer}

To achieve coordination, the nodes need to be deployed as isolated \emph{virtual machines}, forming a fully distributed P2P\arxivfootnote{Peer-to-peer (P2P) computing or networking is a distributed application architecture that partitions tasks or work loads between service peers. Peers are equally privileged participants in the application, and are said to form a peer-to-peer network of nodes \cite{schollmeier2001definition}.} network that can provide support for distributed identity, trust, and transactions.

\paragraph{Virtual Machines (VMs)}
Executing arbitrary code in the machine of a resource-providing user would require a \emph{sandbox}\arxivfootnote{A sandbox is a security mechanism for safely running programs, often used to execute untested code, or untrusted programs from unverified third-parties, suppliers and untrusted users \cite{sandbox}.} for the guest code, a VM\arxivfootnote{A virtual machine is a software implementation of a machine (computer) that executes programs like a real machine \cite{craig2006vm}.} to protect the host. The role of the VM is to make system resources \emph{safely} available to the Community Cloud, upon which Cloud processes could be run safely (without danger to the host machine). Fortunately, feasibility has been proven with heavyweight VMs such as the Java Virtual Machine, lightweight JavaScript VMs present in most modern web browsers, and new approaches such as Google's Native Client. Furthermore, the age \cite{geer2005cmt} of \emph{multi-core processors}\arxivfootnote{A multi-core processor is an integrated circuit to which two or more processors have been attached for enhanced performance, reduced power consumption, and more efficient simultaneous processing of multiple tasks \cite{zelkowitz2007}.} has resulted in unused and underutilised cores being commonplace in modern personal computers \cite{posey2007}, which lend themselves well to the deployment and background execution of Community Cloud facing VMs. Regarding deployment, users would be required to maintain an active browser window or tab, or install a dedicated application. While the first would not require installation privileges, the later would with the benefit of greater functionality. However, more likely a hybrid of both would occur, facilitating the availability and advantages of each in different scenarios.

\paragraph{Distributed Identity}
In distributed systems with variable node reliability, historical context is logically required to have certainty of node interactions. Fundamental to this context is the ability to identify nodes and therefore reference previous interactions. However, current identification schemes have \emph{identity providers} controlling provision. Such as in the DNS\arxivfootnote{The Domain Name System (DNS) is a hierarchical naming-space for computers, services, and other resources participating in the Internet. It translates \emph{domain names} meaningful to humans into their counterpart numerical identifiers associated with networking equipment to locate and address these devices world-wide \cite{mockapetris1988development}. So, translating human-friendly computer hostnames into Internet Protocol (IP) addresses, e.g. www.example.com translates to 208.77.188.166.}, which while nominally distributed, remains under centralised control both technologically and organisationally, permitting numerous distortions in the network. Including \emph{domain squatting}\arxivfootnote{Domain squatting (also known as cybersquatting) is registering, trafficking in, or using a domain name in bad faith, with the intent to profit from the goodwill of a trademark belonging to someone else. The cybersquatter then offers to sell the domain to the person or company who owns a trademark contained within the name at an inflated price \cite{maury2002commerce}.}, abuses by domain registrars \cite{nodaddy.com}, subjection to political control \cite{jonPatel, pakistan-china-DNSSplittingProblems} and risks to the infrastructure being compromised \cite{recentDNSVulnSaga}. Identity in the Community Cloud has to arise naturally from the structure of the network, based on the relation of nodes to each other, so that it can scale and expand without centralised control. We can utilise the property that a large enough identifier-space is unlikely to suffer collisions. For example, the Git distributed version control system \cite{git} assigns a universal identifier to each new commission, without coordination with other repositories. Analogously, assuming each node independently produces a private-public key pair, the probability of public key collision is negligible. Also, from the human identification of nodes we can utilise the property that each node, despite formal identity, possesses a unique position in the network, i.e. set of connections to other nodes. So, combining these two properties provides reasonable certainty for a \emph{distributed identity model} where universal identification can be accomplished without centralised mediation, but this is still an active area of research.

\paragraph{Networking}
At this level, nodes should be interconnected to form a P2P network. Engineered to provide high resilience while avoiding single points of control and failure, which would make decentralised super-peer based control mechanisms \cite{risson2006srt} insufficient. Newer P2P designs \cite{razavi2008sfb} offer sufficient guarantees of distribution, immunity to super-peer failure, and resistance to enforced control. For example, in the Distributed Virtual Super-Peer (DVSP) model a collection of peers logically combine to form a virtual super-peer \cite{razavi2008sfb}, which dynamically changes over time to facilitate fluctuating demands.

\paragraph{Distributed Transactions}
A key element of distributed coordination is the ability of nodes to jointly participate in transactions that influence their individual state. Appropriately annotated business processes can be executed over a distributed network with a transactional model maintaining the ACID\arxivfootnote{ACID (Atomicity, Consistency, Isolation, Durability) is a set of properties that guarantee transactions are processed reliably \cite{haerder1983principles}.} properties on behalf of the initiator \cite{fox1997cbs}. Newer transaction models maintain these properties while increasing efficiency and concurrency. Other directions of research include relaxing these properties to maximise concurrency \cite{vogel}. Others still, focus on distributing the coordination of transactions \cite{razavi2008sfb}. A feature vital for \ac{C3}, as distributed transaction capabilities are fundamental to permitting multi-party service composition without centralised mediation.

\subsubsection{Resource layer}
With the networking infrastructure now in place, we can consider the first consumer-facing uses for the \emph{virtual data centre} of the Community Cloud. Offering the usage experience of Cloud Computing on the \ac{PaaS} layer and above, because Cloud Computing is about using resources from the Cloud. So, Utility Computing scenarios \cite{rappa2004ubm}, such as access to raw storage and computation, should be available at the \ac{PaaS} layer. Access to these abstract resources for service deployment would then provide the \ac{SaaS} layer.

\paragraph{Distributed Computation}
The field has a successful history of centrally controlled incarnations \cite{attiya2004distributed}. However, \ac{C3} should also take inspiration from Grid Computing and Digital Ecosystems to provide distributed coordination of the computational capabilities that nodes offer to the Community Cloud.

\paragraph{Distributed Persistence} 
The Community Cloud would naturally require storage on its participating nodes, taking advantage of the ever-increasing surplus on most\arxivfootnote{The only exception is the recent arrival of \acp{SSD}, popular for mobile devices because of their lack of moving parts, growing in use as their size and price reach that of traditional \acp{HDD} \cite{ssd}.} personal computers \cite{diskTrend}. However, the method of information storage in the Community Cloud is an issue with multiple aspects. First, information can be file-based or structured. Second, while constant and instant availability can be crucial, there are scenarios in which recall times can be relaxed. Such varying requirements call for a combination of approaches, including distributed storage \cite{yianilos2001efd}, distributed databases \cite{garciamolina2008dsc} and key-value stores \cite{keyvaluestore}. Information privacy in the Community Cloud should be provided by the encryption of user information when on remote nodes, only being unencrypted when accessed by the user, allowing for the secure and distributed storage of information.

\paragraph{Bandwidth Management} 
The Community Cloud would probably require more bandwidth at the user nodes than vendor Clouds, but can take advantage of the ever-increasing bandwidth and deployment of broadband \cite{wallsten2008}. Also, P2P protocols such as BitTorrent\arxivOnly{ \cite{cohen2003ibr}} make the distribution of information over networks much less bandwidth-intensive for content providers, accomplished by using the downloading peers as repeaters of the information they receive. \ac{C3} should adopt such approaches to ensure the efficient use of available network bandwidth, avoiding fluctuations and sudden rises in demand (e.g. the Slashdot effect\arxivfootnote{The Slashdot effect, also known as slashdotting, is the phenomenon of a popular website linking to a smaller site, causing the smaller site to slow down or even temporarily close due to the increased traffic \cite{adler1999sea}.}) burdening parts of the network.

\paragraph{Community Currency}

An important theme in the Community Cloud is that of nodes being contributors as well as consumers, which would require a \emph{community currency} (redeemable against resources in the community) to reward users for offering resources \cite{turner2004lightweight}. This would also allow traditional Cloud vendors to participate by offering their resources to the Community Cloud to gather considerable \emph{community currency}, which they can then monetise against participants running a \emph{community currency} deficit (i.e. contributing less then they consume). The relative cost of resources (storage, computation, bandwidth) should fluctuate based on market demand, not least because of the impracticality of predicting or hard-coding such ratios. So, a node of the network would gather \emph{community currency} by performing tasks for the community, which its user could then use to access resources of the Community Cloud.

\paragraph{Resource Repository}
Given that each node providing resources has a different location in the network and quality characteristics, a distributed resource repository would be required that could respond to queries for resources according to desired performance profiles. Such a query would have to consider historical performance, current availability, projected cost and geographical distribution of the nodes to be returned. A constraint optimisation problem, the results returned would be a set of nodes that fit the required profile, proportionally to the availability of suitable nodes.

\subsubsection{Service Layer}
Cloud Computing represents a new era for service-oriented architectures, making services explicitly dependent on other resource providers instead of building on self-sufficient resource locations. \ac{C3} makes this more explicit, breaking down the stand-alone service paradigm, with any service by default being composed of resources contributed by multiple participants. So, the following sections define the core infrastructural services that the Community Cloud would need to provide.

\paragraph{Distributed Service Repository (DSR)}
The service repository of the Community Cloud must provide persistence, as with traditional service repositories \cite{papazoglou2003soc}, for the pointers to services and their semantic descriptions. To support the absence of service-producing nodes during service execution, there must also be persistence of the executable code of services. Naturally, the implementation of a distributed service repository is made easier by the availability of the distributed storage infrastructure of the Community Cloud.

\paragraph{Service Deployment and Execution} 

When a service is required, but is not currently instantiated on a suitable node, a copy should be retrieved from the DSR and instantiated as necessary, allowing for flexible responsiveness and resilience to unpredictable traffic spikes. As nodes are opportunistically interested in executing services to gather \emph{community currency} for their users, so developers should note the resource cost of their services in their descriptions, allowing for pre-execution resource budgeting, and post-execution \emph{community currency} payments. Being in a developer's own interest to mark resource costs correctly, because over-budgeting would burden their users and under-budgeting would cause premature service termination. Additionally, developers could add a subsidy to promote their services. Remote service execution would need to be secured against potentially compromised nodes, perhaps through encrypted processing schemes \cite{1536440}. Otherwise, such nodes while unable to access a complete traffic log of the services they execute, could potentially access the business logic; and we would be replacing the vendor introspection problem, with an anyone introspection problem. Since delivering a service over large distances in the network comes at a potentially high cost, the lack of a central well-connected server calls for a fundamental paradigm shift, from \emph{pull}-oriented approaches to hybrid \emph{push}/\emph{pull}-oriented approaches. So, instead of the \emph{pull}-oriented approach of supplying services only upon request \cite{singh2005soc}, service provision should also follow a \emph{push}-oriented approach of preemptive deployment to strategically suitable nodes, including modifying their deployment profile based on the traffic patterns they face at run time.

\paragraph{Programming Paradigm}

A key innovation of Cloud Computing in its \ac{PaaS} incarnation, is the offering of a well-specified context (programming paradigm) within which the services should be executed \cite{buyya2008market}. The programming paradigm that produces these services is also important to \ac{C3}, because it forms a contract between the service developers and resource providers. The current state-of-the-art requires manipulation of source code in which each line is context dependent, and so a single intended change may necessitates significant alterations at different locations in the codebase. A paradigm shift to \emph{declarative generative programming} \cite{AlexDEST-ArXiv} would be greatly beneficial, avoiding the need to manually manage cascading changes to the codebase. As the requirements behind a service would be made explicit and executable, and being human readable could therefore be manipulated directly as stand-alone artifacts. Additionally, barriers to service composition would be significantly decreased \cite{AlexDESTServComp-ArXiv}, beneficial to \ac{C3} and beyond.

\subsection{Distributed Innovation}
When considering the Community Cloud over time, current software distribution models would cause problems. Should the infrastructure be dependent on a single provider for updates, they would become a single point of control, and possibly failure. Entrusting a single provider with the power to control the evolution of the architecture, even if they are considered benevolent, risks the development goals becoming misaligned with the community. Therefore, the Community Cloud should follow an \emph{evolutionary software distribution model}. Extending an already-growing trend of using distributed code repositories such as \arxivOnly{Git \cite{git} and }Mercurial \cite{mercurial}, over centralised code repositories such as Subversion \cite{svn}\arxivOnly{ and CVS \cite{cvs}}. So, modifications to services, including infrastructural ones, should be distributed locally to migrate over the Community Cloud from where they are deployed, making use of the existing relationships between users. Users or their nodes (by default) could even choose to follow the updates that other trusted peers adopt. Therefore, new versions of a service would compete with older versions, and where superior (fitter) would distribute more widely, spreading further across the Community Cloud. So, updates to services would permeate through the network, in a distributed but regulated manner. We could even consider the updates to services, as the release of patches (modifications), allowing for frequent, smaller and iterative releases more akin to an \emph{evolutionary} software distribution model. Potential speciation (branching) would encourage developers to coordinate their releases and ensure their patches are viable across different branches. Obviously, the ability to undo patches and step back through versions of infrastructural services would be necessary to maintain the Community Cloud. Still, without a more granular approach to conflict resolution from different patching sources, poor developer relations could risk fragmentation of the codebase and network. So, an alternative non-centralised \emph{software innovation model} would be required, such as the \emph{declarative generative programming paradigm} \cite{AlexDEST-ArXiv} mentioned.

\section{In The Community Cloud}

While we have covered the fundamental motivations and architecture of the Community Cloud, its practical application may still be unclear. So, this section discusses the cases of Wikipedia and YouTube, where the application of \ac{C3} would yield significant benefits, because they have unstable funding models, require increasing scalability, and are community oriented.

\subsection{Wikipedia}

Wikipedia suffers from an ever-increasing demand for resources and bandwidth, without a stable supporting revenue source \cite{heebie}. Their current funding model requires continuous monetary donations for the maintenance and expansion of their infrastructure \cite{regwikimon}. The alternative being contentious advertising revenues \cite{heebie}, which caused a long-standing conflict within their community \cite{reuters}. While it would provide a more scalable funding model, some fear it would compromise the content and/or the public trust in the content \cite{wikifear}. Alternatively, the Community Cloud could provide a self-sustaining scalable resource provision model, without risk of compromising the content or public trust in the content, because it would be compatible with their communal nature (unlike their current \emph{data centre} model), with their user base accomplishing the resource provision they require.

Were Wikipedia to adopt \ac{C3}, it would be distributed throughout the Community Cloud alongside other services. With the core operations of Wikipedia, providing webpages and executing server-side scripts, being handled as service requests. Participants would use their \emph{community currency} to interact with Wikipedia, performing a search or retrieving a page, while gaining \emph{community currency} for helping to host Wikipedia across the Community Cloud. More complicated tasks, such as editing a Wikipedia webpage, would require an update to the distributed storage of the Community Cloud, achieved by transmitting the new data through its network of nodes, most likely using an eventual consistency model \cite{vogel}.

\subsection{YouTube}

YouTube requires a significant bandwidth for content distribution, significant computational resources for video transcoding, and is yet to settle on a profitable business model \cite{utubemon2, youtubeMoney}. In the Community Cloud, websites like YouTube would also have a self-sustaining scalable resource provision model, which would significantly reduce the income required for them to turn a profit.

Were YouTube to adopt \ac{C3}, it would also be distributed throughout the Community Cloud alongside other services. Updates such as commenting on a YouTube video, would similarly need to propagate through the distributed persistence layer. So, the community would provide the bandwidth for content distribution, and the computational resources for video transcoding, required for YouTube's service. The \ac{QoS} requirements for YouTube are significantly different to those of Wikipedia, because while constant throughput is desirable for video streaming, occasional packet loss is tolerable. Also, YouTube's streaming of live events has necessitated the services of bespoke content distribution networks \cite{utubeAkamai}, a type of service for which the Community Cloud would naturally excel.

We have discussed Wikipedia and YouTube in the Community Cloud, but other sites such as arXiv and Facebook would equally benefit. As \ac{C3}'s organisational model for resource provision moves the cost of service provision to the user base, effectively creating a micro-payment scheme, which would dramatically lower the barrier of entry for innovative start-ups.

\section{Conclusions}

We have presented the Community Cloud as an alternative to Cloud Computing, created from blending its usage scenarios with paradigms from Grid Computing, principles from Digital Ecosystems, self-management from Autonomic Computing, and sustainability from Green Computing. So, \ac{C3} utilises the spare resources of networked personal computers to provide the facilities of data centres, such that the community provides the computing power for the Cloud they wish to use. A socio-technical conceptualisation for sustainable distributed computing.

While the Open Cloud Manifesto \cite{openCloudManifesto} is well intentioned, its promotion of open standards for vendor Cloud interoperability has proved difficult \cite{openCloudProblems}. We believe it will continue to prove difficult until a viable alternative, such as \ac{C3}, is developed. Furthermore, we hope that the Community Cloud will encourage innovation in vendor Clouds, forming a relationship analogous to the creative tension between open source and proprietary software.

In the future we will continue to refine the various elements of \ac{C3}, such as suitable mechanisms for a \emph{community currency}, distributed alternatives to DNS, DVSPs, RESTful Clouds, \emph{declarative generative programming paradigms}, distributed innovation, and the environmental impact of the Community Cloud relative to vendor Clouds.

\section*{Acknowledgements}
We would like to thank for comments and helpful discussions Paulo Siqueira of the Instituto de Pesquisas em Tecnologia e Inovacao, Eva Tallaksen and Alexander Deriziotis.

\bibliographystyle{IEEEtran.bst}
\nocite{hewitt2008osr, pauloGreenComp, weiss2007cc, c3}
\bibliography{references,cloudReferences}

\end{document}